# Thermal diffuse scattering: elastic and coherent

Yuan Yao

Beijing National Laboratory of Condensed Matter Physics, Institute of Physics, Chinese Academy of Sciences, Beijing 100190, People's Republic of China

Corresponding authors: yaoyuan@iphy.ac.cn

Abstract

Thermal diffuse scattering (TDS) caused by the interaction of high-energy electrons with phonons has been investigated. An oscillating atom retains all of its elastic scattering capacity, although the vibration changes the spatial distribution of the scattering potential. Therefore, the scattering by the vibrating lattice or phonons is still elastic. Furthermore, the TDS electrons remain coherent and form the diffuse interference patterns even under the time-averaging assumption. Elastic and coherent TDS electrons contribute most of the contrast of Kikuchi lines.

1. Introduction

Thermal diffuse scattering (TDS) is a historical topic in the scattering physics of transmission electron microscopy (TEM). Similar to X-ray diffraction, this concept was introduced to account for the decay of the Bragg reflection intensity. TDS results from the thermal vibration of the atoms, which breaks the periodic arrangement of the atoms, weakening the diffraction strength and producing a diffuse scattering background. Yoshioka[1] proposed a general theory that considered the inelastic scattering process. He derived a coefficient representing the transfer of energy from incident electrons to vibrating atoms based on Bloch waves and attributed the intensity loss of Bragg spots to this inelastic scattering. Later, Takagi[2] pointed out that the difference $\Delta V(\vec{r})$ between the scattering potential $V(\vec{r})$ of an atom and its time-average $<V(\vec{r})>$ should be responsible for the diffuse scattering. This difference can be taken as an absorption potential that violates the Bloch theorem and reduces the intensity of the Bragg reflection. Since $\Delta V(\vec{r})$ depends on how the atom deviates from its equilibrium, the loss of scattering potential is related to the phonon vibration modes. This assumption directly related the atomic scattering potential to the thermal fluctuation, so that $\Delta V(r)$ or phonon vibration spectra can be formally applied to calculate diffuse scattering using the perturbation transition matrix of Bloch waves. Although such a treatment did not regard $\Delta V(r)$ as the energy transfer from incident electrons to phonons, Takagi argued that the diffuse scattering accompanies the emission or absorption of the phonons, implying that the scattering is inelastic and $\Delta V(r)$ should be equivalent to the phono absorption or a decaying potential. Yoshioka and Kainuma[3] found that the calculated intensity based on the inelastic phonon scattering was much smaller than the attenuation of Bragg scattering, but they still introduced an absorption potential originating from the lattice fluctuation to describe the thermal influence on diffraction. Whelan[4] explicitly included the phonon scattering into the inelastic scattering theory to calculate the attenuation of Bragg reflections. He used the phonon absorption potentials defined by Takagi and related it to Debye-Waller (DW) factor. Einstein model was chosen to simplify the phonon vibration model. So an inelastic framework that links the absorption potential (imaginary potential) and the DW factor has been established. It is an energy transfer process and therefore incoherent scattering. Rez et al[5] extended it to involve the partial coherence effect.

However, Hall and Hirsch[6] insisted that elastic scattering from vibrating atoms dominates the diffuse feature. They claimed that TDS is essentially elastic scattering, although phonon interactions modify the scattering potential. They then employed an effective absorption potential and Einstein model to predict the fading of Bragg reflections and interpreted the diffuse scattering as the coherent superposition of the Bloch waves with the DW factor. Here, the absorption potential derived from the DW factor or Einstein model is only a tool for expecting the attenuation of Bragg spots and does not imply a true energy transfer between incident electrons and phonons. This view actually laid the theoretical foundation for simulating the diffuse intensity by superimposing elastic scattering waves that are generated from various atom configurations constructed by the frozen lattice (FL) or frozen phonon (FP) model in the multislice algorithm.

Another important issue is whether the electrons scattered by phonons are coherent. In principle, inelastically scattered electrons should be incoherent in most cases while elastically scattered electrons may or may not be coherent, depending on the specific scattering process. Van Dyck[7] studied the coherence of inelastically scattered electrons and concluded that TDS should be incoherent for its inelastic nature. Wang[8] regarded TDS as "quasi-elastic" scattering and proved that the diffuse scattering is incoherent based on the inelastic theory of Yoshioka, even with Einstein model. He also barrowed Takagi's scheme that the instantaneous scattering potential $V(\vec{r})$ of an atom is greater than the time-averaged static potential $<V(\vec{r})>$ to illuminate the contrast of TDS.[9] Ten years later Van Dyck[10] tried to resolve the contradiction between the elastic nature in the multislice program with FL model and the inelastic property of the phonon absorption. He demonstrated that the integrated diffuse intensity from various elastic configurations with quantum-mechanical treatment should in fact be equivalent to the inelastic distribution generated by the phonon vibration, with the implicit assumption that $\Delta V(\vec{r})$ correlates the energy transfer to the lattice vibration. Van Dyck's statement ensured the adequacy of FL or FP model in image simulation without inelastic mechanism or absorption potential, and the quantitative interpretation has been achieved in the experimental high angle annular dark field (HAADF) images where TDS dominates the contrast.[11]

Here, the physical basis of TDS is revisited and the elastic and coherent characteristics are favored.

2. Theory
2.1 Elastic or inelastic?

High-energy incident electrons can induce many inelastic scattering processes, such as plasmon oscillation, atom excitation and phonon absorption, etc. Plasmon oscillations and shell transitions rarely occur spontaneously at room temperature due to their higher exciting energy, but phonon vibrations are always present because of thermal fluctuation in the environment. An essential fact is that the population of phonons at a given temperature is not determined by the incident electrons although high energy electrons may produce additional phonons within the sample. The DW factor is a measure of the intrinsic phonon distribution at ambient temperature, and does not represent the state of electron-excited phonons. Thus it is not correct to estimate the energy absorbed by the excited phonon using the DW factor because to do so means that all phonons in the sample are produced by electron bombardment, which obviously exaggerates the electron-phonon interaction. In fact, the temperature increase due to incident beam is limited and locally confined. Thus, the calculations using phonon population or the DW factor are not feasible even for true inelastic phonon scattering.

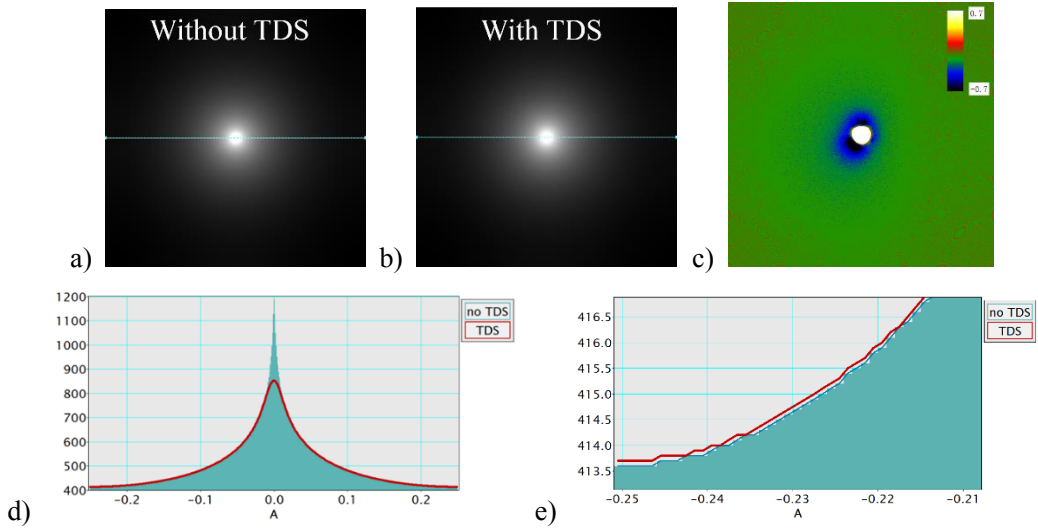

Fig. 1 a) Project potential V($\vec{r}$) of a Si atom. b) Average project potential <V($\vec{r}$)> of a vibrating Si atom. c) ΔV($\vec{r}$) = V($\vec{r}$) - <V($\vec{r}$)>. d) Comparison of the line profiles in a) and b). d) Magnification of the border of d). The red line is the data with the thermal fluctuation effect.

The discrepancy ΔV($\vec{r}$) between the scattering potential V($\vec{r}$) of an atom and its time-averaged static <V($\vec{r}$)> is real. Takagi and his followers considered it as a real loss of scattering capability and implemented it directly as an inelastic perturbation to predict the TDS intensity. However, this assumption is questionable since the atom vibration will definitely redistribute the elastic scattering potential in real space, not increase or decrease it. To investigate the effect of atom vibration on the scattering potential, the spatial distribution of the project scattering potential of a Si atom was simulated with and without thermal oscillation (Fig. 1a and Fig. 1b). The DW factor was 0.5 and 100 positions following Gaussian random distribution were utilized to simulate the thermal oscillation. Fig. 1c and Fig. 1d compare the difference between two potential maps and the corresponding line profiles, respectively. At first glance, it is apparent that ΔV($\vec{r}$) is strongly positive around the equilibrium, indicating that atomic vibration dilutes the scattering potential. This feature confirms Takagi's view and similar images calculated with and without the DW factor were also shown by Wang[9] to conclude that the decaying potential due to thermal vibration represents the loss of scattering capability. However, the total intensity of 52.49 in Fig. 1c is almost zero compared to the total intensity of $1.093×10^8$ in Fig. 1a or Fig. 1b. Fig. 1c and Fig. 1e show that the time-averaged potential has a larger value outside the central region. Although the intensity of each pixel does not exceed much, the number of peripheral pixels compensates for the loss of potential in the center. This property is not captured by the potential map of Wang because a decaying potential with the DW factor there failed to portray the excess part away from the center. Indeed, it is easy to recognize that the vibrating atom can spread the incoming electrons over a wider area, sacrificing the central strength. ΔV($\vec{r}$) is positive around the equilibrium position but negative elsewhere. So the total scattering potential is not lost as expected, it is just redistributed in real space! A reasonable conclusion from the conservation of the scattering potential is that the scattering from the vibrating atom is still elastic. In fact, the electron energy loss spectrum (EELS) experiment testified the overwhelming superiority of elastically scattered electrons over inelastically scattered electrons in diffuse diffraction.[12] It convinces the validity of the elastic scattering assumption in the multislice method with FL or FP model when calculating TDS intensity.

2.2 Spatial coherence or incoherence?

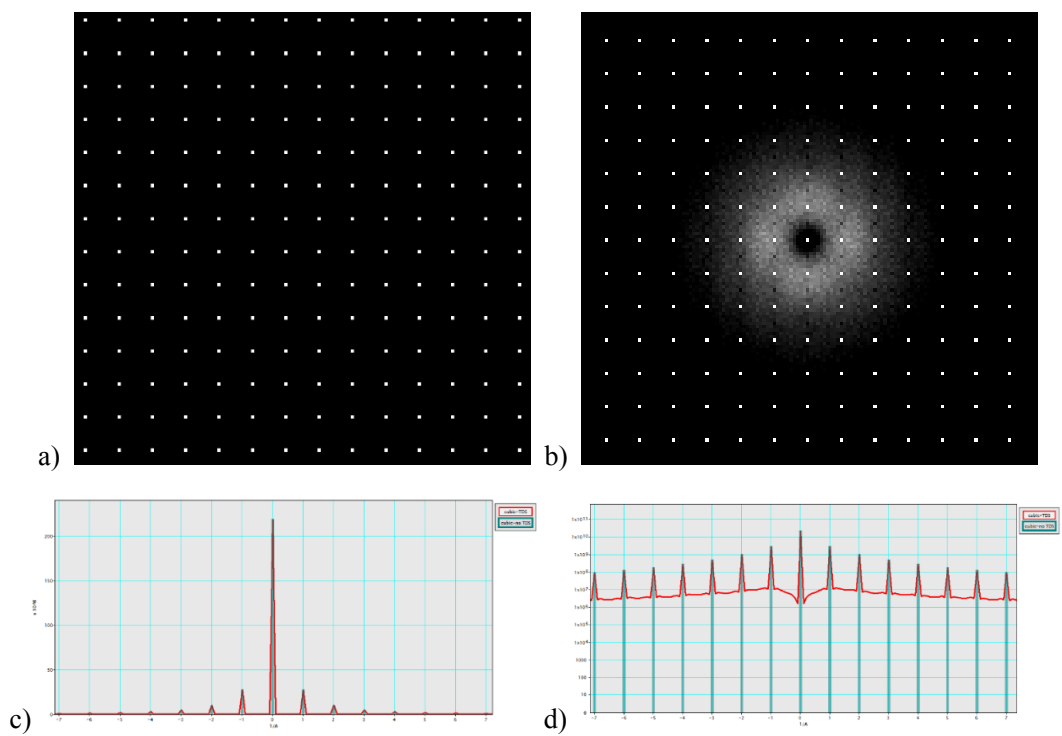

Fig. 2 Simulated diffraction of artificial cubic structure containing Si atoms with static configuration a) and with FL model b), respectively. Comparison of the diffraction intensity of static and thermal vibration diffraction by linear scale c) and logarithmic scale d), respectively. The red line is the profile of diffraction intensity containing TDS.

Fig. 2a and Fig. 2b show the simulated diffraction patterns for an artificial crystal with and without thermal diffuse effect, respectively. Si atoms were arranged on a 10 × 10 cubic lattice with a space of 3 Å. To simplify the calculation, the project potential was generated from one layer of atoms with a sampling of 0.01 Å. The absolute value of Fast Fourier Transform (FFT) of the project potential was considered as the diffraction intensity. FL model with the DW factor of 0.5 was implemented and diffractions from 1000 configurations were averaged to achieve TDS feature. Obviously, the FL model can reproduce the diffuse pattern in the diffraction image (Fig. 2b).

An important point is that the diffuse feature is not naturally associated with incoherence. It is a well-known phenomenon that a disordered structure can give rise to a coherent diffuse feature, such as the broad diffuse ring formed by amorphous carbon. Therefore, it is wrong to distinguish between coherent and incoherent diffraction based on whether a scattered electron falls on the Bragg point or not. Unfortunately, the mistake of directly judging TDS as the incoherent scattering just because of its diffuse feature has appeared in some literature.

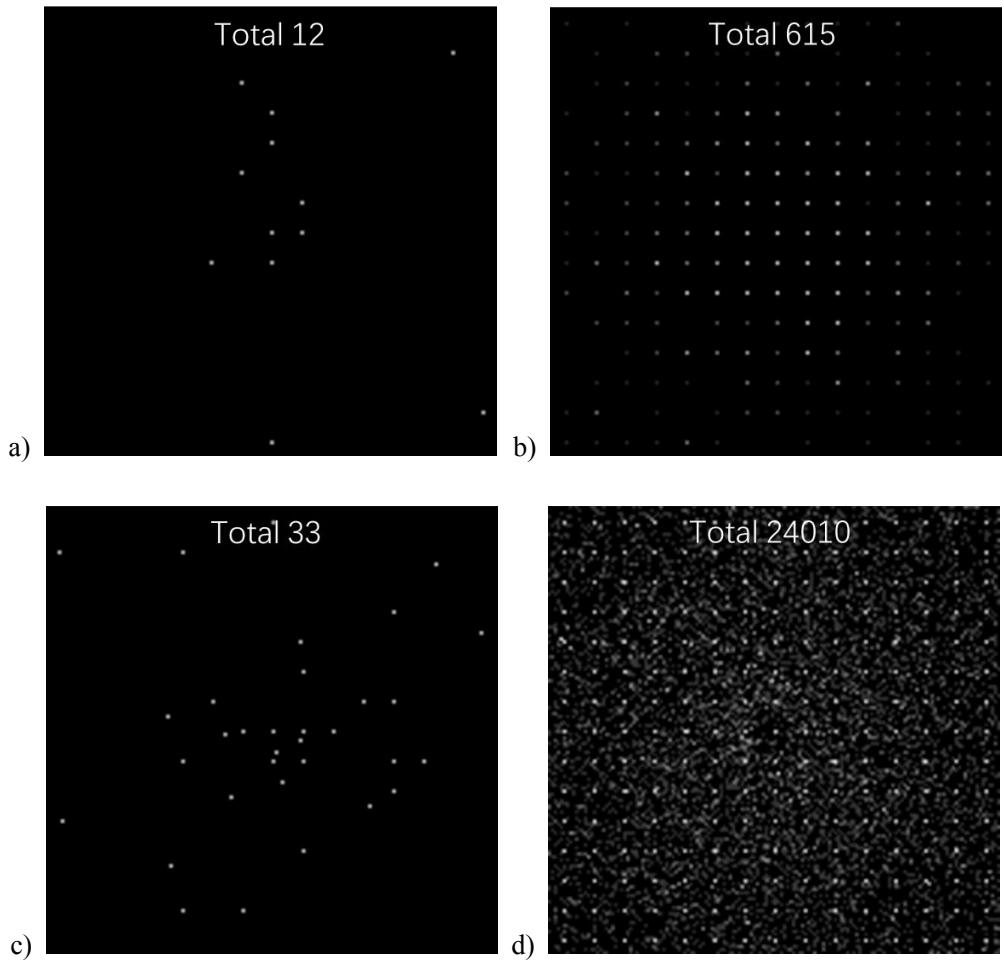

Fig. 3 Simulated diffraction patterns for independent electron events. a) and b) for static structure, c) and d) for vibrating lattice. The total number of electrons is shown.

To elucidate the coherent or incoherent behavior of thermally scattered electrons, virtual "single electron scattering" experiments were performed by simulation. The intensity of a pixel $(i,j)$ on the Fourier Transform (FT) of the exit wave was converted into the probability $p(i,j)$ of an electron falling on this pixel. Electrons arrived at the imaging plane one by one, and the chance of registering at a particular pixel depended on its $p(i,j)$. Higher $p(i,j)$ meant that more electrons appeared there. So it is explicitly equivalent to integrating the snapshot of single electron scattering. Each snapshot in reciprocal space ($k$-space) captured the coherent scattering characteristic. Thousands of snapshots were synthesized into the diffraction features, just like the time-averaging process. The integration of snapshots is certainly an incoherent buildup. Fig. 3a is the result of a static atom configuration with a small amount of electron scattering, which is undoubtedly elastic and coherent scattering. However, without a priori knowledge of the sample structure, it is impossible to distinguish which point is the Bragg reflection because there is no regular distribution. Actually, all points in the $k$-space are equal, no particular point naturally correlates to a Bragg reflection or coherent scattering. As more electrons are acquired, the Bragg features gradually appear (Fig. 3b). From a quantum mechanical perspective, each electron also carries the probability of coherent exit wave to the imaging plane. "Incoherent summation" of the coherent diffraction of singe electron still displays the property of coherence and produces the correct Bragg reflections in Fig. 3b. Single-electron interference experiment achieved by Tonomura et al. attested that the Young's interference fringes appeared after long-exposure imaging of the independent electrons, verifying that time-averaging does not destroy the interference of scattering.[13]

Now a similar procedure was repeated in TDS simulation with the FL model. It is important to emphasize that each snapshot was a FT of an array of elastic scattering potentials carrying a thermal fluctuation. Thus each snapshot still mirrors the "coherent interference" of the elastically scattered electrons by the vibrating elastic scattering potentials corresponding to a particular lattice configuration. It is difficult to distinguish the Bragg reflections in the low-dose image (Fig. 3c) while sharp spots emerge after sufficient electron accumulation (Fig. 3d), accompanied by a diffuse background. As mentioned above, all points are equal in $k$-space and one should not specify that one corresponds to a coherent or incoherent scattering. The diffuse intensity beyond the regular spots also reflects the "coherent characteristic" of the electrons scattered by the vibrating lattice as well, precisely due to the thermal effect, although it is the fruit of "incoherent summation" or time-averaging.

3. Discussion

High-energy electrons can certainly excite extra lattice vibrations or phonons in the characterized sample. The creation of new phonons is the true energy absorption for the incident beam and leads to the inelastic scattering. However, the contribution of this inelastic event to the TDS is small compared to the elastic scattering by the vibrating atoms. Using the phonon population, e. g. at room temperature, to estimate the inelastic scattering strength of phonons with perturbation theory is wrong, because in reality the electron beam does not evoke that many phonons. The assumption that a vibrating atom reduces its elastic scattering capability is also incorrect. The elastic potential preserves its scattering capability so that the inelastic treatment with the energy loss $\Delta V(\vec{r})$ or the DW factor should be abandoned.

The perturbation approach in the elastic framework proposed by Hall and Hirsch is a valid way to solve the scattering attenuation of Bragg reflections by recombining the elastic Bloch waves with the DW factor.[6] The "absorption potential" for the calculation of the TDS intensity should be used with caution, since it is a virtual concept used to describe the dilution of Bragg spots and does not correspond to a real energy transfer in the usual sense. The multislice method, including the FL or FP model, is practical for implementing TDS simulation because it depicts the correct physical picture of thermal elastic scattering and the coherence characteristics. The DW factor and the thermal vibration behind it do not weaken the scattering ability, but break the structure order and thus cause the diffuse feature.

In principle, the elastically scattered electrons from the vibrating atoms are coherent to forge the interference patterns. The interference patterns are not always sharp and regular Bragg spots, so the diffuse feature cannot simply be treated as a consequence of incoherent effects. The contrast of diffraction in real experiments or in virtual single-electron-scattering experiments also demonstrates the coherence characteristic of the elastically scattered electrons by the phonons. Bragg reflections are the coherent interference of scattered electrons when the atoms are in their equilibrium position, while TDS reflects the coherent interference of the scattered electrons when the atoms deviate from their equilibrium position. Statistically, the probability of all atoms locating in the equilibrium position is very small, but this configuration can result in a strong in-phase consequence which greatly promotes the chance of electrons appearing in Bragg spots. The out-of-phase electrons scattered by the disordered atoms should be spread over a larger area and form weak contrast because of the low probability of appearance in these "non-Bragg" pixels, although these configurations are high probability events. From the perspective of signal processing, different configurations contain common frequency components corresponding to the regular lattice, so that superposition of the FT results in the enhancement of Bragg reflections. This may be the reason why the Bragg reflections in the real diffraction patterns are very sharp and bright.

The inelastic scattering associated with energy absorption by phonons actually occurs as electrons penetrate the sample and contributes to the diffuse contrast. The proportion of this inelastic scattering is small when examining the EELS shape involving phonon vibration in the literature.[12] Discussion of scattering associated with phonon absorption is beyond the scope of this paper. How many phonons are excited by incident electrons or what the phonon spectrum looks like in relation to the energy exchange between incident electrons and phonon interactions requires careful evaluation.

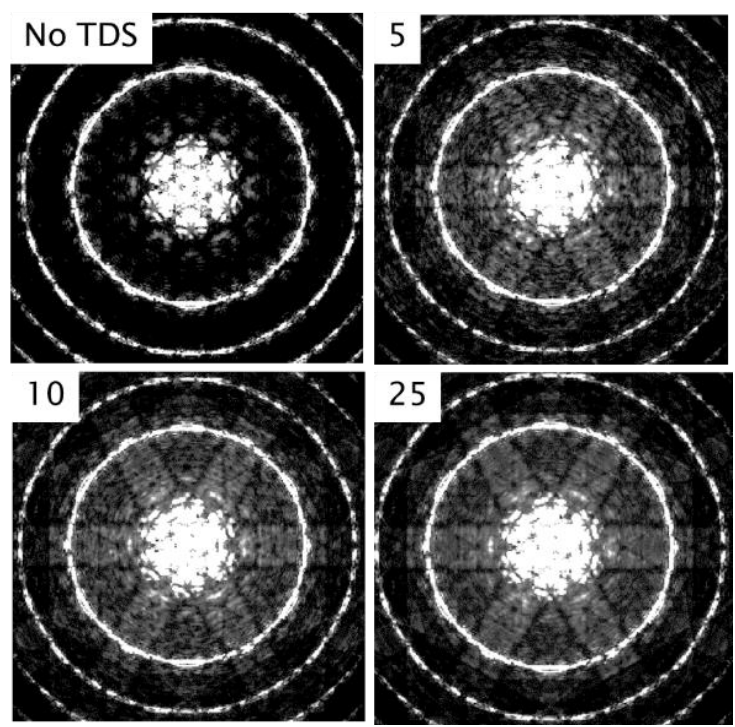

Fig. 4 Simulated CBED patterns of [111] Si. The number of configurations is shown. Accelerating energy: 100 kV, convergence angle: 8 mrad, Cs: 3.3 mm, defocus: -140 nm, sampling: 0.05 Å/pixel, image size: 1024 × 1024, thickness: 489 Å.

Another interesting phenomenon related to TDS is Kikuchi lines. The multislice simulations of convergent beam electron diffraction (CBED) with and without FL model for [111] Si are shown in Fig. 4. As the number of configurations increases, Kikuchi lines become stronger. It at least proves that elastic potential and Einstein model can reproduce the Kikuchi patterns. Some literature also confirmed this result.[12, 14]

Note that inelastically scattered electrons can also form the coherent interference, similar to the single-electron diffraction experiments described above. However, due to the slight difference in electron wavelength caused by the energy dispersion, the diffraction generated by inelastically scattered electrons is more diffuse when the thermal influence is taken into account too. Multislice simulation with Einstein model cannot mimic the contribution of real phonon absorption to TDS unless a dispersion effect is introduced, so any method that uses iterative algorithms to quantitatively match experimental image contrast in ptychography should consider this minor deficiency.[15, 16]

Conclusion

The summary is simple. Thermal vibration modulates the spatial distribution of the elastic scattering potential of atoms but the total scattering capability is still preserved. TDS is dominated by the oscillating elastic scattering potential. The diffuse feature of TDS reflects the coherent interference of the electrons scattered by the fluctuating lattice, despite of the time-averaging assumption.

Acknowledgement

This work was supported the National Key R&D Program of China (No. 2022YFB3803900).